\documentstyle[11pt,newpasp,twoside,epsf]{article}
\def\edcomment#1{\iffalse\marginpar{\raggedright\sl#1\/}\else\relax\fi}
\reversemarginpar

\begin{document}
\title{The Galactic Double-Neutron-Star Merger Rate: Most Current Estimates}
 \author{C.\ Kim$^1$, V.\ Kalogera$^1$, D.R.\ Lorimer$^2$, M.\ Ihm$^1$, and K.\ Belczynski$^{1,3}$}
\affil{$^{1}$Northwestern University, Department of Physics and Astronomy, 2145 Sheridan Rd., Evanston, IL, 60201, USA}
\affil{$^{2}$University of Manchester, Jodrell Bank Observatory, Macclesfield,
Cheshire, SK11 9DL, UK}
\affil{$^{3}$Lindheimer Postdoctoral Fellow}

\begin{abstract}
We summarize our results on the Galactic merger rate of double neutron
stars (DNS) in view of the recent discovery of PSR J0737$-$3039. We
also present previously unpublished results for the {\em global\/}
probability distribution of merger rate values that incorporate the
presently known systematics from the radio pulsar luminosity
function. The most likely value obtained from the global distribution
is only $\simeq15\,{\rm Myr}^{-1}$, but a re-analysis of the current
pulsar sample and radio luminosities is needed for a reliable
assessment of the best fitting distribution. Finally, we use our
theoretical understanding of DNS formation to calculate a possible
upper limit on the DNS merger rate from current Type Ib/c supernova
rate estimates.
\end{abstract}

\section{Introduction}

Soon after the discovery of the highly relativistic pulsar
J0737$-$3039 (Burgay et al.\ 2004) we applied our analysis method for
pulsar populations and calculated the updated merger rate estimates
for the current sample of Galactic close DNS (Kalogera et al.\
2004). Our main conclusion was that this new, remarkably relativistic
system is very significant for these estimates and leads to a rate
increase by a factor of $5-7$. This implies a correspondingly
significant increase in DNS inspiral event rates for
gravitational-wave (GW) interferometers like LIGO. In what follows, we
summarize our recent results and present new results on: (i) how a
{\em global\/} probability distribution of rate estimates can be
calculated with the systematic uncertainties; (ii) the possible upper
limits that could be imposed on the DNS merger rate if we adopt the
theoretically expected relationship between DNS merger rates and Type
Ib/c supernova (SN) rates.

\section{The revised Galactic DNS merger rate}

In Kim, Kalogera, \& Lorimer (2003; hereafter KKL), we introduced a
statistical method to calculate the probability density function (PDF)
of the rate estimates for Galactic close DNS. After the discovery of
PSR~J0737$-$3039, we derived a combined $P({\cal R})$ considering the
three observed DNS systems in the Galactic disk (for details see
Appendix~A of Kim et al.\ 2004).

To calculate the merger rate of DNS systems in our Galaxy, we need to
estimate: (i) the number $N_{\rm tot}$ of Galactic pulsars with pulse
and orbital characteristics {\it similar\/} to those in the observed
sample; (ii) the lifetime $\tau_{\rm life}$ of each observed system;
(iii) an upward correction factor $f_{\rm b}$ for pulsar beaming.

We calculate $N_{\rm tot}$ by modeling in detail the pulsar-survey
selection effects for a number of pulsar population models described
in KKL. The model assumptions for the pulsar luminosity function
dominate the systematic uncertainties of our overall calculation.

The lifetime of the system is defined by $\tau_{\rm life} \equiv
\tau_{\rm sd} + \tau_{\rm mrg}$, where $\tau_{\rm sd}$ is a spin-down
age of a recycled pulsar (Arzoumanian, Cordes, \& Wasserman 1999) and
$\tau_{\rm mrg}$ is the remaining lifetime until the two neutron stars
merge (Peters \& Mathews 1963). We note that the lifetime of
J0737$-$3039 is estimated to be 185\,Myr, which is the shortest among
the observed systems.

The beaming correction factor $f_{\rm b}$ is defined as the inverse of
the fractional solid angle subtended by the pulsar beam. Its
calculation requires detailed geometrical information on the
beam. Following Kalogera et al.\ (2001), we adopt $f_{\rm b}=5.72$ for
PSR B1913+16 (Hulse \& Taylor 1975) and 6.45 for PSR B1534+12
(Wolszczan 1991). Without good knowledge of the geometry of
J0737$-$3039A, we adopt the average value of the other two systems
($\simeq 6.1$).

In Figure~1, we show $P({\cal R})$ for our chosen reference model that
allows for a low minimum pulsar luminosity (Model~6 in KKL). The most
likely value of ${\cal R}$ turns out to be $83\,{\rm Myr}^{-1}$,
larger by a factor of $\simeq6.4$ than the rate estimated before the
discovery of J0737$-$3039. We find the same increase factor for all
pulsar population models examined. This revised merger rate implies an
increase in the detection rate of DNS inspirals for ground-based GW
interferometers such as LIGO (Abramovici et al. 1992). Using the
standard extrapolation of our reference model out to extragalactic
distances (see Kalogera et al.\ 2001), we find that the most probable
event rates are 1 per 29\,yrs and 1 per 2 days, for initial and
advanced LIGO, respectively. At the 95\% confidence interval, the most
optimistic predictions for the reference model are 1 event per 8\,yrs
and 2 events per day for initial and advanced LIGO, respectively. For
more details see Kalogera et al.\ (2004). 

\begin{figure}
\plotone{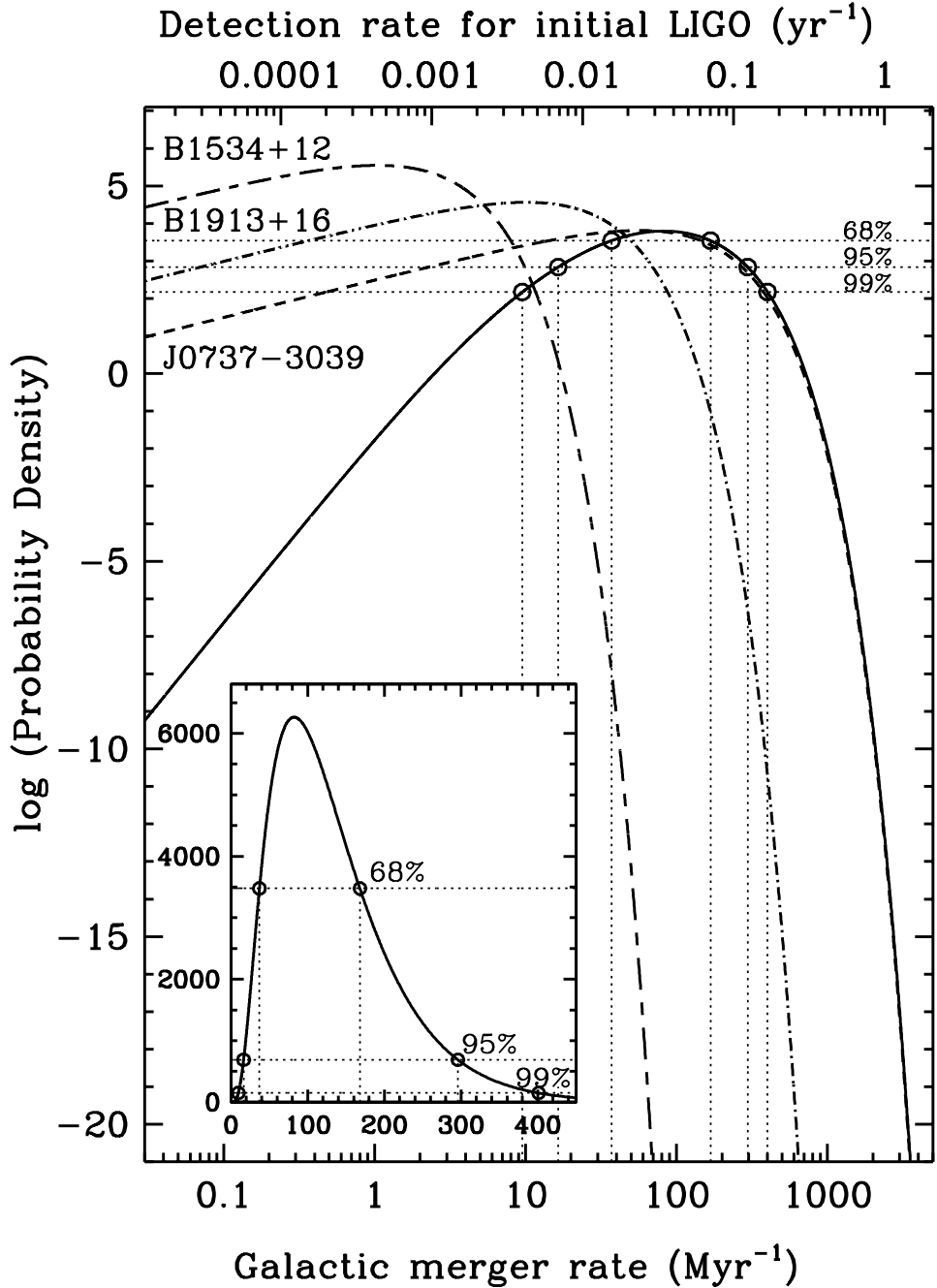}
\caption{The PDF of DNS merger rate $P({\cal R})$ is shown on a log
scale.  The thick solid line is the total Galactic rate estimate
overlapped with results for individual observed systems (dashed
lines).  Dotted lines indicate confidence intervals for the rate
estimates. The same results are shown on a linear scale in the small
inset. All results shown are for our reference model.}
\end{figure}

The revised DNS merger rate is dominated by PSR
J0737$-$3039. Therefore, if the estimated lifetime of this system is
revised in the future, it will directly affect our rate
estimation. Lorimer et al.\ (in this volume) calculated the spin-down
age of the system with various spin-down models and suggested an age
in the range 30--70\,Myr, which is shorter than the value we adopted
for our calculation ($\tau_{\rm sd}=$100\,Myr). The edges of this
range give us rate estimates of ${\cal R}\simeq 90-115\,$Myr$^{-1}$.

The beaming correction for J0737$-$3039 is also important for our rate
estimation. Recently, Jenet \& Ransom (2004) suggested a geometrical
model for this newly discovered system. According to their model, the
predicted beaming correction factor is $\geq$6 
assuming a two-sided beam (Jenet 2004, private communication). 
If confirmed, this could lead to a further dramatic increase in the merger rate estimates.

\section{Global probability distribution of the rate estimates}

In KKL, we showed that estimated Galactic DNS merger rates are
strongly dependent on the assumed luminosity distribution function for
pulsars. So far, we have reported results for each set of population
model assumptions. Here we describe how we can incorporate the
systematic uncertainties from these models and calculate, $P_{\rm g}
({\cal R})$, a {\em global\/} PDF of rate estimates. However, we
stress that the information needed for such a calculation is currently
not up to date; therefore, specific quantitative results could
change when constraints on the luminosity function are derived from
the current pulsar sample.

We calculate $P_{\rm g}({\cal R})$ using the prior distributions of
the two model parameters for the pulsar luminosity function: the
cut-off luminosity $L_{\rm min}$ and power-index $p$. We calculate
these priors by fitting the marginal PDFs of $L_{\rm min}$ and $p$
presented by Cordes \& Chernoff (1997). We obtain the following
analytic formulae for $f(L_{\rm min})$ and $g(p)$: $f(L_{\rm min}) =
\alpha_{\rm 0} + \alpha_{\rm 1} L_{\rm min} + \alpha_{\rm 2} L_{\rm
min}^{2}$ and $g(p) = 10^{\beta_{\rm 0} + \beta_{\rm 1} p + \beta_{\rm
2} p^{2}}$, where $\alpha_{\rm i}$ and $\beta_{\rm i} ~({\rm
i}=0,1,2)$ are coefficients we obtain from the least-square fits and
the functions are defined over the intervals $L_{\rm min}=[0.0,\,1.7]$
mJy kpc$^{2}$ and $p=[1.4,\, 2.6]$. We note that, although Cordes and
Chernoff (1997) obtained $f(L_{\rm min})$ over $L_{\rm min}\simeq
[0.3,\,2]$ mJy kpc$^{2}$ centered at 1.1 mJy kpc$^{2}$, we consider
$f(L_{\rm min})$ with a peak at $\sim 0.8\,$mJy kpc$^{2}$ considering
the discoveries of faint pulsars with L$_{\rm 1400}$ below 1 mJy
kpc$^{2}$ (Camilo 2003).

We use the above priors to calculate $P_{\rm g}({\cal R})$:
\begin{equation}
P_{\rm g}({\cal R}) = \int_{p} dp \int_{L_{\rm min}} ~dL_{\rm min} P(R) f(L_{\rm min}) g(p)~.
\end{equation}
In Figure~2, we show the distributions of $L_{\rm min}$ and $p$
adopted (top panels) and the resulting global distribution of Galactic
DNS merger rate estimates (bottom panel). We find that $P_{\rm
g}({\cal R})$ is strongly peaked at {\em only\/} around
15\,Myr$^{-1}$. We note that this is a factor $\simeq$ 5.5 smaller
than the revised rate from the reference model (${\cal
R}=$83\,Myr$^{-1}$). At the 95\% confidence interval, we find that
the Galactic DNS merger rates lie in the range $\sim$
1--170\,Myr$^{-1}$. These imply LIGO event rates in the range 
$\sim (0.4-70)\times 10^{-3}\,$yr$^{-1}$ (initial) and 
$\sim 2-380\,$yr$^{-1}$ (advanced). 
Given these implications, it is clear
that up-to-date constraints on $L_{\rm min}$ and $p$ and their PDFs (a
follow-up on Cordes \& Chernoff 1997) are urgently needed.

\section{Rate constraints from Type Ib/c supernovae and binary 
evolution models}

Based on our current understanding of DNS formation, the progenitor of
the second neutron star is expected to form during a Type Ib/c
supernova. Therefore, the empirical estimates for the Type Ib/c SN
rate in our Galaxy can be used to provide upper limits on the DNS
merger rate estimates. From Cappellaro, Evans, \& Turatto (1999) we
adopt ${\cal R}_{\rm SN\,Ib/c} \simeq 1100\pm500\,$Myr$^{-1}$ 
(for Sbc--Sd galaxies). Here, we assume $H_0=71\,$km/s/Mpc and 
$L_{\rm B,gal}=9\times10^{9}\,L_{\rm B,sun}$.

\begin{figure}
\plotone{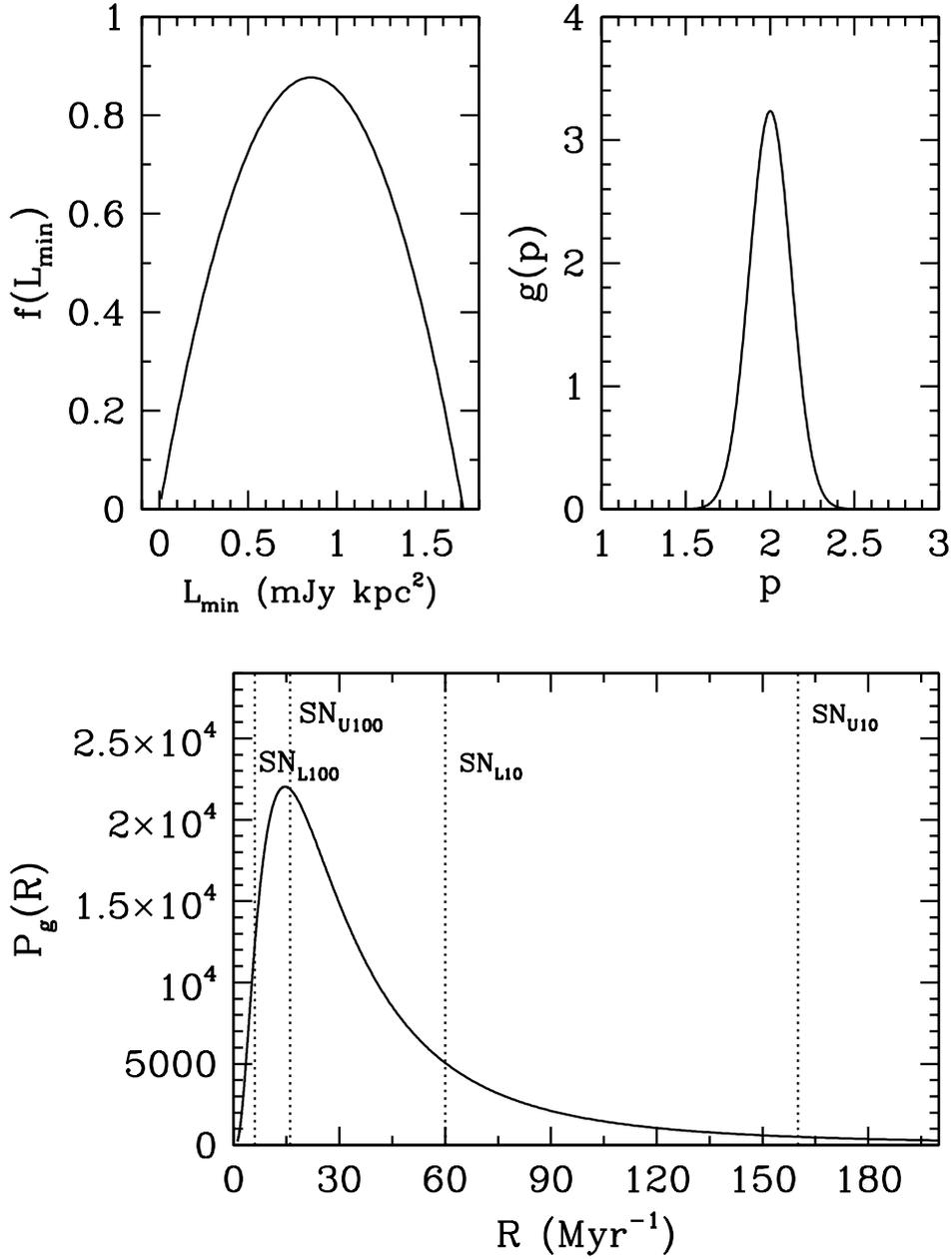}
\caption{The global $P_{\rm g}({\cal R})$ on a linear scale (lower
panel) and the assumed intrinsic distributions for $L_{\rm min}$ and
$p$ (upper panels). Dotted lines represent the lower ($SN_{\rm L}$)
and upper ($SN_{\rm U}$) bounds on the observed SN Ib/c rate scaled by
1/10 and 1/100 (see text). The empirical SN Ib/c rates range over
$\sim 600-1600\,$Myr$^{-1}$, where the average is at $\sim$1100
Myr$^{-1}$ (Cappellaro, Evans, \& Turatto 1999), beyond the range
shown here.}
\end{figure}

In order to find the fraction of SN Ib/c actually involved in the
formation of DNS, we have examined population synthesis models
calculated with the code {\tt StarTrack} (Belczynski, Kalogera, \&
Bulik 2002; Belczynski et al.\ 2004) and find very low rate ratios:
$\gamma\equiv{\cal R}$/${\cal R}_{\rm
SN\,Ib/c}\sim0.001-0.005$. Several models with He-star winds
consistent with current observations (weaker than previously thought)
lead to $\gamma\simeq 0.005$. We note that systematic overestimation
of ${\cal R}_{\rm SN\,Ib/c}$ relative to SN~II rates has already been
pointed out (Belczynski, Kalogera, \& Bulik 2002; this is related to
the assumption of {\em complete\/} removal of H-rich
envelopes). However, we find that this discrepancy would raise the
value of $\gamma$ by just a factor of a few. As an approximate
constraint, we adopt the empirical ${\cal R}_{\rm SN\,Ib/c}$ and scale
it by 1/10 and 1/100, reflecting the results from population synthesis
calculations. We overlay these scaled values in Fig. 2 (dotted lines
in the bottom panel) using the ranges for SN Ib/c reported by
Cappellaro, Evans, \& Turatto (1999).

We note that our most optimistic DNS merger rate is 
${\cal R}=224^{+594}_{-181}$ Myr$^{-1}$ at a 95\% confidence interval 
(Model~15 in KKL). We obtain $\gamma$ for SN Type Ib/c to be $\sim$0.8 with the upper limit of ${\cal R}$ at the 95\% confidence interval. This corresponds to $\gamma \sim$0.1 with a SN Type II rate, which is factor 6.1 larger than that of SN Type Ib/c. In both cases, the most optimistic model is lower than the current empirical supernova rate estimates, but not really consistent with the results of population synthesis calculations. If we consider the global distribution, with the upper limit of ${\cal R}$ at the 95\% confidence interval, we obtain $\gamma$ $\sim$0.15 and 0.025 for SN Type Ib/c and II, respectively.

\section{Prediction for more DNS detections by the PMB survey}

Acceleration searches of the PMB-survey data (Faulkner et al.\ 2003)
should significantly improve the detection efficiency of DNS
binaries. Although the data analysis is on-going, acceleration
searches already led to the discovery of PSR J1756$-$2251 (see
Faulkner et al.\ 2004 and the contribition by Lyne in this volume).

Following the method described in Kalogera, Kim \& Lorimer (2003), we
calculate the probability to detect a pulsar similar to any of the
observed DNS systems. We assume that acceleration searches can
perfectly correct for the reduction in flux due to Doppler smearing,
namely no degradation in the calculation of signal-to-noise ratio for
the PMB survey is included. Considering observed DNS systems
individually, we calculate the expected number of pulsars to be
detected by the PMB survey ($N_{\rm exp}$).

The probability distribution of $N_{\rm exp}^i$ for each DNS pulsar
sub-population $i$ (B1913+16, B1534+12, J0737--3039) is given by:
\begin{equation}
 P_i(N_{\rm exp})={\frac{{\beta_i}^{2}}{(1+\beta_i)^{2}}} 
{\frac{({N_{\rm exp}}+1)}{(1+{\beta_i})^{N_{\rm exp}}} }~, 
\end{equation}
where the constants $\beta_i$ are a measure of how less likely it is
to detect pulsars without acceleration searches relative to with
acceleration searches. For each sub-population, we calculate the mean
values of $N_{\rm exp}$, which are $\bar N_{\rm exp, 1913} = 0.9$,
$\bar N_{\rm exp, 1534} = 1.2$, and $\bar N_{\rm exp, 0737} = 1.9$.

\begin{figure}
\plotfiddle{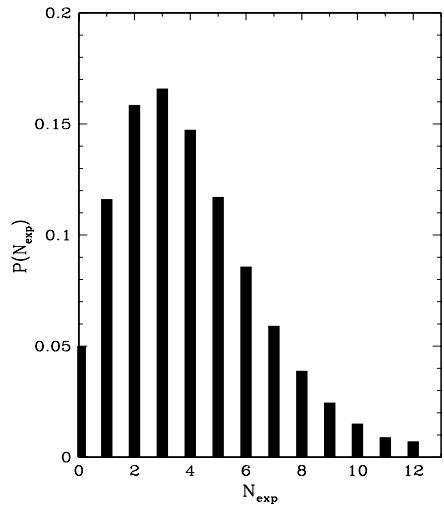}{6.4cm}{0}{30}{26}{-100}{-15}
\caption{$P(N_{\rm exp})$ for close DNS systems in the PMB survey.
We consider all three observed systems in the Galactic disk. The mean value
is $\bar N_{\rm exp}=$4.0. The result shown here is based on our reference model.}
\end{figure}

The increase of the observed sample is very important for the
reduction of the uncertainties associated with the DNS merger rate
estimates. We note, however, that the discovery of new systems that
are {\em similar\/} to the three already known does not necessarily
imply a significant increase in the rate estimates. Significant
changes are expected when new systems are discovered with pulse
profiles or binary properties significantly different than the old
ones, as those systems will reveal a new DNS sub-population in the Galaxy.

\vskip 20pt

\acknowledgments
We would like to thank Kip Thorne for suggesting incorporating the
systematics into a single PDF, Takashi Nakamura and Steinn Sigurdsson
for raising the question of Ib/c SN rates, and Frederick Jenet and
Thomas A. Prince for useful discussions. This research is partially
supported by NSF Grant 0121420, and a Packard Foundation Fellowship in
Science and Engineering to VK. DRL is a University Research Fellow
supported by the Royal Society. He also thanks the Theoretical
Astrophysics Group at Northwestern University for support. KB is a
Lindheimer Fellow at Northwestern University and also acknowledges
support from grant PBZ-KBN-054/p03/2001.

\end{document}